# Lag synchronization and scaling of chaotic attractor in coupled system

Sourav K. Bhowmick[1], Pinaki Pal[2], Prodyot K. Roy[3], Syamal K. Dana[1]

[1]*Central Instrumentation, CSIR-Indian Institute of Chemical Biology, Kolkata* 700032*, India*
[2]*Department of Mathematics, National Institute of Technology, Durgapur* 713209*, India*
[3]*Department of Physics, Presidency University, Kolkata* 700073*, India*

**Abstract:** We report a design of delay coupling for lag synchronization in two unidirectionally coupled chaotic oscillators. A delay term is introduced in the definition of the coupling to target any desired lag between the driver and the response. The stability of the lag synchronization is ensured by using the Hurwitz matrix stability. We are able to scale up or down the size of a driver attractor at a response system in presence of a lag. This allows compensating the attenuation of the amplitude of a signal during transmission through a delay line. The delay coupling is illustrated with numerical examples of 3D systems, the Hindmarsh-Rose neuron model, the Rössler system and a Sprott system and, a 4D system. We implemented the coupling in electronic circuit to realize any desired lag synchronization in chaotic oscillators and scaling of attractors.

PACS numbers: 05.45.Gg, 05.45.Xt

**Lead Paragraph**

**Lag synchronization (LS) is usually seen in mismatched chaotic oscillators for instantaneous coupling in the weaker limit. Alternatively, LS can be realized in chaotic oscillators by introducing a delay in the linear coupling and this delay can be varied within a stability limit. In practical applications, particularly, during a transmission of signal through a delay line, attenuation of the signal is a common problem. A transmitted signal needs to be amplified for exact retrieval of the signal at the other end of the transmission line. On the other hand, amplification (or attenuation) of a chaotic signal using linear active devices is challenging due to bandwidth limitation of such devices. We address the issue of LS with a scaling (amplification or attenuation) of chaotic signal, in this paper, using a design of coupling approach. The coupling function is assumed unknown *a priori*. Given the description of a dynamical system, the coupling is defined using an open-plus-closed-loop (OPCL) scheme that ensures asymptotic stability of LS with a delay in the coupling. Once the coupling is designed and switched on between the systems, the interacting oscillators synchronize asymptotically. The concept of coupling strength does not exist in the definition. The issue of amplification or attenuation of a chaotic signal during a transmission is, particularly, given a priority. A scaling parameter is purposefully introduced in the definition of the delay coupling which can be varied smoothly for realization of an exact amplified (or attenuated) version of a chaotic driver attractor at a response system keeping the LS state undisturbed.**

## I. INTRODUCTION

Lag synchronization appears[1-4] in mismatched chaotic oscillators under linear instantaneous coupling in the weaker coupling limit. In this LS state, the amplitudes of the similar pairs of state variables of the coupled oscillators remain correlated but shifted in time. This is in contrast to the phase synchronization (PS)[1] when the amplitudes lose correlation but maintained a constant phase difference. LS onsets[2,3,5] in mismatched oscillator above a critical coupling larger than the critical coupling for PS but lower than the critical coupling for complete synchronization (CS) in identical oscillators. The amount of time lag between the instantaneously coupled oscillators depends upon the amount of mismatch and the coupling strength. Practically, a tradeoff between the mismatch and the coupling strength is necessary to decide the time lag. The mismatch parameter can be detuned[5] either positively or negatively to induce lag or lead synchronization. But one can control the delay or lag time within a limit. Beyond this limit, the amplitudes of the oscillators, however, start losing correlation[4]. Alternatively, LS is also implemented[6] in coupled chaotic systems by introducing an explicit time delay in the coupling function where the lag time can be varied arbitrarily. The stability of the LS depends upon the coupling strength[6-11] and system parameters. However, the amplitude distortion cannot be avoided during the transmission of a chaotic signal through a long transmission line which actually decides the lag time. Note that other distortions also exist but the amplitude distortion has a dominant effect. A delayed replica of a chaotic signal at a distant oscillator coupled by a delay line is desired for practical applications. For this we need to amplify the delayed version of the original signal (or the driver signal) at a far end. Linear amplifying devices are available for efficient amplification of a periodic signal but its gain is limited by the frequency bandwidth. It restricts the use of a linear amplifier for a broadband chaotic signal. A method of projective lag synchronization[12] has been proposed which is able to scale up or down the size of an attractor. However, no attempt is made, to our best knowledge, for practical implementation of such projective lag synchronization.

In this paper, we propose a design of coupling that allows physical realization of the LS in chaotic oscillators. We realize LS in two unidirectionally delay coupled oscillators where any desired time lag or delay $\tau$, (even larger than the characteristic time scale of the oscillator) can be introduced to a chaotic system. In contrast to the existing linear coupling, the proposed method is able to induce LS in both identical and mismatched oscillators. The stability of the LS is robust to mismatch in the coupled oscillators which is an advantage in a practical situation. Especially, we focus on the scaling of the size of a chaotic attractor at a response state. For this, we introduce a parameter in the coupling definition which can be smoothly controlled to amplify or even to attenuate a chaotic signal. We mainly extend the existing theory of open-plus-closed-loop (OPCL)[13] for instantaneous coupling to delay coupling using the Hurwitz matrix stability to ensure stability of the LS scenario. We first elucidate the OPCL method with numerical examples of identical bursting Hindmarsh-Rose neuron model[14], and of mismatched 3-dimensional system such as Rössler and Sprott systems[15] and, of a 4-dimensional system. Finally, we implement the coupling scheme in electronic circuit.

The paper is organized as follows: the theory of time-delayed OPCL coupling is described in section II. A numerical example of Hindmarsh-Rose model is presented in section IIIA to elaborate the LS in identical oscillators. Next we described the mismatched case in section IIIB with numerical examples of a Rossler system and a 4D system in section IIIC. Finally, numerical result of a Sprott system with experimental verification is described in section IV. Results are concluded in section V.

## II. TIME-DELAYED OPCL COUPLING

We earlier reported[13] an instantaneous unidirectional OPCL coupling for realization of complete synchronization (CS), antisynchronization (AS) with a scaling (amplification or attenuation) of chaotic signal, generalized synchronization[16] in two chaotic oscillators and AS in identical complex networks[17]. Here we extend the OPCL method to unidirectional time-delayed coupling and realize LS in two chaotic oscillators, identical or mismatched. For this, we make a modification in the OPCL coupling to include a delay in the coupling definition. Most importantly, we introduce a constant parameter in the coupling function which can be easily varied to scale up or down the size of the attractor of a chaotic driver at a response system.

A chaotic driver is described by $\dot{y} = f(y) + \Delta f(y)$, $y \in R^n$, where $\Delta f(y)$ contains the mismatch terms. The response is described by $\dot{x} = f(x), x \in R^n$. Now, we define a goal dynamics[13] by $g_\tau = \alpha y_\tau$ where $y_\tau = y(t-\tau) = x(t)$ is a desired response, $\tau$ is a lag time and $\alpha$ is an arbitrary multiplicative constant. We mention that, instead of taking $\alpha$ as a constant, we may replace this by a constant matrix which will lead to generalized lag synchronization (GLS), i.e., generalized synchronization[17] with an induced lag time. Presently, we consider the simpler case of LS where $\alpha$ is a constant. Hence the response system with a delay coupling is given by

$$\dot{x} = f(x) + D(x, g_\tau) \qquad (1)$$

where the delay coupling term $D(x, g_\tau)$ is defined by

$$D(x, g_\tau) = \dot{g}_\tau - f(g_\tau) + \left(H - \frac{\partial f(g_\tau)}{\partial g_\tau}\right)(x - \alpha y_\tau). \qquad (2)$$

$\frac{\partial f(g_\tau)}{\partial g_\tau}$ is the *Jacobian* of the system and $H$ is an arbitrary constant matrix ($n \times n$) as usual[13]. Using a Taylor's series expansion of $f(x)$ and restricting to the first order derivative, the error dynamics, $e = (x - \alpha y_\tau)$, can be easily derived from (1) and (2) as $\dot{e} = He$. Now, if the eigenvalues of $H$ all have negative real parts (such a matrix is called a Hurwitz matrix) for appropriate choice of the elements of the $H$-matrix, the error dynamics ensures that $e \to 0$ as $t \to \infty$ and we obtain asymptotically stable LS. By choosing $\alpha = 1$, the response variables become identical to those of the driver but shifted in time by the desired time $\tau$. The success of the design of OPCL coupling depends upon the choice of the elements of the $H$-matrix. To realize any desired LS, the most important task is to construct the $H$-matrix in (2) and to convert it to a Hurwitz by appropriate choice of its elements. Instead of wildly searching for an appropriate $H$-matrix, it is noticed[13] that $H$ can be easily constructed from the *Jacobian* of a given dynamical system by applying a set of thumb rules: if the $ij$th element, $J_{ij}$, of the *Jacobian* is a constant, we keep it as the $ij$th element of the $H$-matrix, i.e., we choose $H_{ij}=J_{ij}$. Otherwise if the $ij$th element involves a state variable, we choose $H_{ij}=p_k$ where $p_k$'s (k=1, 2, ..) stands for new constant parameters. This leads to a simple way of constructing the OPCL delay coupler. All the parameters of the $H$-matrix are assumed defined once the system is known with its parameters except the constant parameter $p_k$ that is to satisfy a Routh-Hurwitz (RH) criterion[18]. The RH criterion guarantees the $H$-matrix to be a Hurwitz. For a $3 \times 3$ matrix, the characteristic equation is $\lambda^3 + a_1\lambda^2 + a_2\lambda + a_3 = 0$ and the RH criterion is satisfied if $a_1>0$, $a_3>0$, $a_1a_2>a_3$ and this determines the constant $p_i$. For 4D systems, the characteristic equation is a polynomial of degree 4: $a_4\lambda^4 + a_3\lambda^3 + a_2\lambda^2 + a_1\lambda + a_0 = 0$. The R-H condition is for all $a_i>0$ and, $a_3a_2>a_4a_1$, $a_3a_2a_1>a_4a_1^2 + a_3^2a_0$. Although this makes the design of the coupling more complicated but is still manageable. This completes the design of the coupling and we elaborate this with examples in the next sections.

It is not difficult to implement the method once the Hurwitz matrix is constructed from the *Jacobian* of the interacting oscillators. We are able to control the amount of lag time ($\tau$) between a driver and a response oscillator at any desired or preset value. Furthermore, the constant $\alpha$ in the goal dynamics can be used as a control parameter for amplification ($|\alpha|>1$) or attenuation ($0<|\alpha|<1$) of the driver attractor. In other words, the concept of projective lag synchronization or the scaling of an attractor at a response system is incorporated in the definition of the present coupling scheme by the use of the multiplying constant $\alpha$. We emphasize that, once the coupling is defined and inserted between the drive and the response, one can smoothly vary $\alpha$ to change the size of the response attractor at a LS state. Furthermore, once LS is attained after the initial transients, the variation of $\alpha$ does not disturb its stability. The same is true for the variation in $\tau$. Thus the major advantage of the proposed method lies in practical realization of LS and, a smooth and precise control over it. The flexibility of controlling a lag time enhances possibility of applications such as encoding information in 2D or 3D arrays of coupled oscillators. The scaling of the size of a chaotic attractor in presence of LS is another major practical advantage.

**IIIA. Numerical Simulation: Identical Oscillators**

We first implement the OPCL delay coupling in two identical oscillators: a bursting Hindmarsh-Rose (HR) neuron model[14] is taken as a driver,

$$\dot{y}_1 = y_2 - ay_1^3 + by_1^2 - y_3 + I,$$
$$\dot{y}_2 = c - dy_1^2 - y_2, \quad (3)$$
$$\dot{y}_3 = r\{S(y_1+1.6) - y_3\}.$$

where $y_1$ is the membrane potential, $y_2$ and $y_3$ are associated with the fast and the slow membrane currents, $I$ is the input bias current to the neuron. The *Jacobian* of the model is

$$\frac{\partial f}{\partial y} = \begin{bmatrix} -3ay_1^2 + 2by_1 & 1 & -1 \\ -2dy_1 & -1 & 0 \\ rS & 0 & -r \end{bmatrix}. \quad (4)$$

The H-matrix is now constructed from the *Jacobian*,

$$H = \begin{bmatrix} p_1 & 1 & -1 \\ p_2 & -1 & 0 \\ rS & 0 & -r \end{bmatrix}. \quad (5)$$

We consider another identical HR system as a response system. The coupling $D(x, g_\tau)$ for the HR model is then derived using (2) and (5). The response HR oscillator after coupling is obtained as,

$$\dot{x}_1 = x_2 - ax_1^3 + bx_1^2 - x_3 + I$$
$$+ a\alpha(\alpha^2 - 1)y_{1\tau}^3 + b\alpha(1-\alpha)y_{1\tau}^2 + (\alpha-1)I$$
$$+ \{p_1 - 2b\alpha\, y_{1\tau} + 3a(\alpha\, y_{1\tau})^2\}(x_1 - \alpha\, y_{1\tau}),$$
$$\dot{x}_2 = c - dx_1^2 - x_2 + c(\alpha-1) + d\alpha\,(\alpha-1)y_{1\tau}^2$$
$$+ (p_2 + 2d\alpha\, y_{1\tau})(x_1 - \alpha\, y_{1\tau}), \quad (6)$$
$$\dot{x}_3 = r\{s(x_1+1.6) - x_3\} + 1.6rs(\alpha-1).$$

where $y_{i\tau} = y_i(t-\tau)$

The parameters ($p_1, p_2$) in the *H*-matrix (5) are now so chosen as to satisfy the RH criterion: $p_1<r+1$ for this HR model if we assume $p_2=0$. For a set of selected parameters, $a = 1.3, b = 3, c = 0.3, d = 5, r = .005, S = 5, I = 5$, we choose $p_1=-2$ to define the coupling for realizing an asymptotically stable LS. We consider $\alpha=1$ when the response system is simplified,

$$\dot{x}_1 = x_2 - ax_1^3 + bx_1^2 - x_3 + I + (p_1 - 2by_{1\tau} + 3a\, y_{1\tau}^2)(x_1 - y_{1\tau}),$$
$$\dot{x}_2 = c - dx_1^2 - x_2 + (p_2 + 2d\, y_{1\tau})(x_1 - y_{1\tau}), \quad \dot{x}_3 = r\{s(x_1+1.6) - x_3\}. \quad (7)$$

Numerical examples for two desired time lags are shown in Fig. 1. Clearly the response variable $x_1(t)$ in dotted lines (online blue) follows the driver variable $y_1(t)$ in solid lines (online red) with a time shift. Time series of the driver $y_1(t)$ and the response $x_1(t)$ are shown for two different lag time, $\tau=4.99$, 14.99 in Fig. 1(a) and 1(c) respectively. It is important to note that all the other state variables too maintain the same time lag. This is an important and unique consequence of the OPCL coupling. The driver and their corresponding delayed version shows coincidence in Fig. 1(b) and (1(d) for $\tau=4.99$, 14.99 respectively. We confirm the LS quantitatively for both the cases using a similarity measure[2,3] ($\sigma$) to determine the time lag between two time series of the driver, $y_1(t)$, and the response, $x_1(t)$,

$$\sigma = \frac{\langle [x_1(t) - y_1(t-\tau_s)]^2 \rangle}{[\langle x_1(t)^2 \rangle \langle y_1(t)^2 \rangle]^{1/2}} \quad (8)$$

The similarity measure, when plotted with time delay $\tau$, shows a global minimum $\sigma=\sigma_{min}=0$ for a delay $\tau_s=\tau_{so}$ that determines the lag time in LS. This confirms the realization of the desired LS in two identical systems, as targeted in two different cases, $\tau=4.99$, 14.99 in Figs. 1(e) and 1(f) respectively. In reality, one can vary $\tau$-value without loss of synchronization and thereby implement LS.

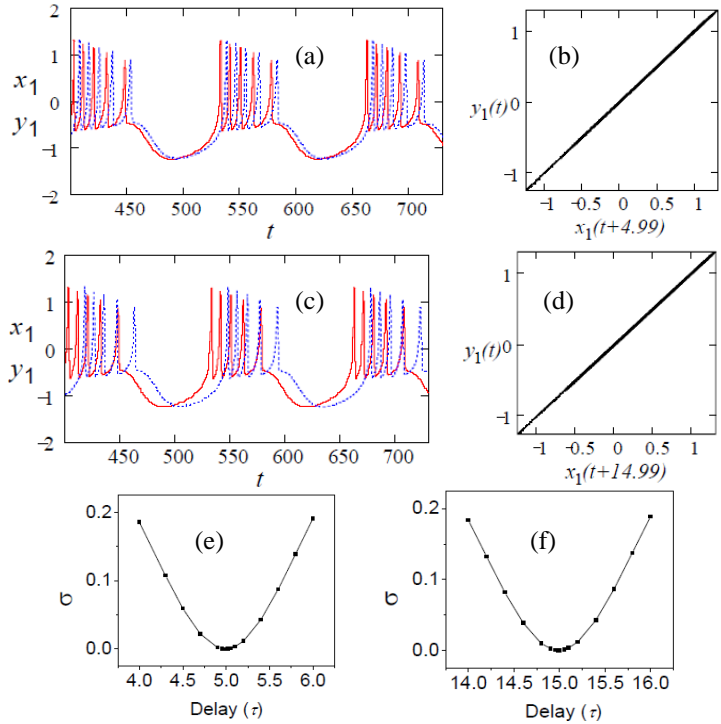

**Fig.1** (color online). Lag synchronization in identical Hindmarsh-Rose neuron model. Time series of two driver variable $y_1$ (red) and response variable $x_1$ (blue) for delays (a) $\tau=4.99$, (c) $\tau=14.99$. The corresponding synchronization manifold in (b) and (d) are shown immediately at their right. The minima of the similarity function plots in (e) and (f) at $\tau=4.99$, $\tau=14.99$ respectively confirm the desired LS. Driver and response parameters: $a = 1.3$, $b = 3$, $c = 0.3$, $d = 5$, $r = .005$, $S = 5$, $I = 5$. Hurwitz parameters, $p_1 = -2$, $p_2 = 0$.

### IIIB. Numerical Examples: Mismatched Rössler Oscillators

In a second example, we show that the LS scenario is valid even in presence of mismatch in the systems: a Rössler system is taken a driver,

$$\dot{y}_1 = -\omega y_2 - y_3; \quad \dot{y}_2 = y_1 + by_2 + \Delta by_2$$
$$\dot{y}_3 = c + y_3(y_1 - d) + \Delta c - \Delta dy_3 \quad (9)$$

where $\omega$, $b$, $c$, $d$ are the system parameters and, $\Delta c$ and $\Delta d$ represent mismatches. The response oscillator with OPCL delay coupling,

$$\dot{x}_1 = -\omega x_2 - x_3; \quad \dot{x}_2 = x_1 + bx_2 + \alpha\Delta b y_{2\tau}$$
$$\dot{x}_3 = c + x_3(x_1 - d) + \alpha\Delta c + \alpha(1-\alpha)y_1 y_3$$
$$+ (p_1 - \alpha y_{3\tau})(x_1 - \alpha y_{1\tau}) + (p_2 - \alpha y_{1\tau})(x_3 - \alpha y_{3\tau})\quad(10)$$

where $y_{i\tau} = y_i(t-\tau)$

Details of constructing the *H*-matrix in Rössler oscillator and making a choice of the parameters of the *H*-matrix to be a Hurwitz, are given in ref.13. We use the same Hurwitz matrix to define the coupling function: the instantaneous coupling term and the goal dynamics is replaced by its delayed version. Numerical simulations show LS of lag $\tau$=3.99 in Fig. 2(a) since we inserted the same delay in the coupling. A lag of $\tau$=3.99 is confirmed by the coincidence of $y_1(t)$ with the time shifted response signal, $x_1(t+3.99)$ in Fig.2(b). Similarly, we can implement any arbitrary LS in the coupled systems. This allows a smooth control of the arbitrary lag time which is an important benefit for engineering applications. Note that we introduce mismatch in more than one parameter and a stable LS scenario is still realized. The proposed design of delayed OPCL coupling overrules, in contrast to the previous reports[6], the effect of parameter mismatch in realization of arbitrary LS in the driver-response system.

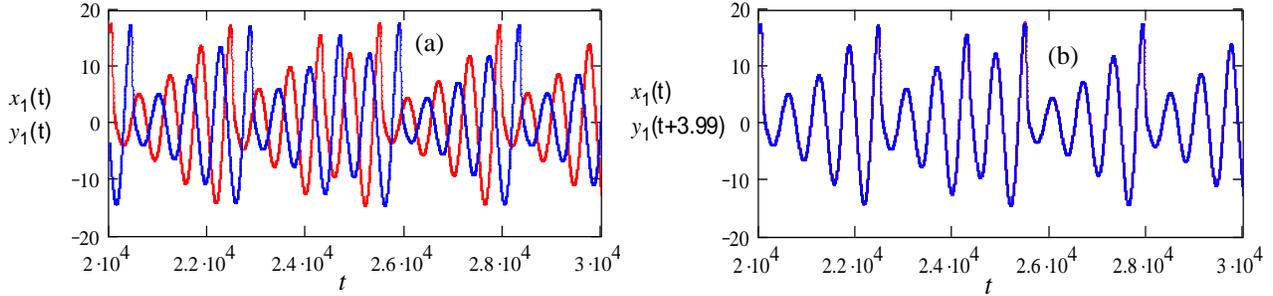

**Fig.2** (color online). LS in mismatched Rössler system. Time series (after the transient) of driver signal $y_1$ (red line) and the response signal $x_1$ (blue line) for time delays $\tau$=3.99 in (a). Time series $y_1(t)$ coincide with time shifted response signal $x_1(t+3.99)$ in (b). Driver parameters: $\omega = 1$, $b = 0.18$, $c = 0.2$, $d = 10$, response parameters: $\omega = 1$, $b_2 = 0.16$, $\Delta c = -0.04$ and $\Delta b = 0.02$, $d = 10$. Hurwitz matrix for the Rössler system, H= [0 -1 -1; 1 $b$ 0; $p_1$ 0 $p_2$]$^T$, where $p_1 = 1.5$, $p_2 = 4$. Scaling parameter $\alpha$=1.

### IIIC. Numerical Examples: 4 dimensional System

Now we now demonstrate the applicability of the coupling design for LS in higher dimensional system. We use a 4D system which is a modified version of the Lorenz system,

$$\dot{y}_1 = \delta(y_2 - y_1),$$
$$\dot{y}_2 = (r + \Delta r)y_1 - y_2 - y_1 y_3,$$
$$\dot{y}_3 = -by_3 + y_1 y_2,$$
$$\dot{y}_4 = -y_1 y_3 + dy_4. \quad(11)$$

where the parameters are chosen as $\delta$=10, $r$=28, $b$=8/3, $d$= -10 for a chaotic oscillation, while one mismatch parameter is introduced as $\Delta r$=0.5. The *Jacobian* and the *H*-matrix of the 4D system (11) are given by,

$$J = \begin{bmatrix} -\delta & \delta & 0 & 0 \\ r+\Delta r - y_3 & -1 & y_1 & 0 \\ y_2 & y_1 & -b & 0 \\ -y_3 & 0 & -y_1 & d \end{bmatrix}; \quad H = \begin{bmatrix} -\delta & \delta & 0 & 0 \\ p_1 & -1 & p_2 & 0 \\ p_3 & p_4 & -b & 0 \\ p_5 & 0 & p_6 & d \end{bmatrix}$$

The *H*-matrix is obtained using the basic guidelines discussed above for 3D systems and it has $p_i$ ($i$=1, 2, 3, 4, 5, 6) parameters. We can make many choices of the $p_i$ parameters; one typical choice is $p_1$= -50, $p_2$=$p_3$=$p_4$=$p_5$=$p_6$=0. The response system after coupling now appears as

$$\begin{aligned}
\dot{x}_1 &= \delta(x_2 - x_1) \\
\dot{x}_2 &= rx_1 - x_1 x_3 - x_2 + \alpha \Delta r y_{1\tau} + (\alpha^2 - \alpha) y_{1\tau} y_{3\tau} \\
&\quad + (p_1 + \alpha y_{3\tau})(x - \alpha y_{1\tau}) + (p_2 + \alpha y_{1\tau})(x_3 - \alpha y_{3\tau}) \\
\dot{x}_3 &= -bx_3 + x_1 x_2 + (\alpha - \alpha^2) y_{1\tau} y_{2\tau} \\
&\quad + (p_3 - \alpha y_{2\tau})(x_1 - \alpha y_{1\tau}) + (p_4 - \alpha y_{1\tau})(x_3 - \alpha y_{2\tau}) \\
\dot{x}_4 &= -x_1 x_3 + dx_4 + (\alpha - \alpha^2) y_{1\tau} y_{3\tau} \\
&\quad + (p_3 - \alpha y_{3\tau})(x_1 - \alpha y_{1\tau}) + (p_2 + \alpha y_{1\tau})(x_3 - \alpha y_{3\tau})
\end{aligned} \qquad (12)$$

We present here only the case of $\alpha=1$ for the 4D system although other values of $\alpha$ are always applicable. The response system is then simplified,

$$\begin{aligned}
\dot{x}_1 &= \delta(x_2 - x_1) \\
\dot{x}_2 &= rx_1 - x_1 x_3 - x_2 + \Delta r y_{1\tau} + (p_1 + y_{3\tau})(x - y_{1\tau}) + (p_2 + y_{1\tau})(x_3 - y_{3\tau}) \\
\dot{x}_3 &= -bx_3 + x_1 x_2 + (p_3 - y_{2\tau})(x_1 - y_{1\tau}) + (p_4 - y_{1\tau})(x_3 - y_{2\tau}) \\
\dot{x}_4 &= -x_1 x_3 + dx_4 + (p_3 - y_{3\tau})(x_1 - y_{1\tau}) + (p_2 + y_{1\tau})(x_3 - y_{3\tau})
\end{aligned} \qquad (13)$$

For numerical simulation, we consider two different lag time, $\tau=0.2, 3.0$ as introduced in the coupling function. A pair of time series of the driver ($y_1$) and the response ($x_1$) in dotted (blue) and solid (red) lines respectively are shown in Fig. 3(a) and 3(b) for two different lag time. The similarity plots of the pair of time series in Fig. 3(a) and 3(b) are drawn in Fig. 3(c) and 3(d) respectively which confirm that they are exactly realized as desired. Two cases of lag time, one smaller and another larger are shown to evidence that the lag time can be chosen arbitrarily even to extent of the characteristic time scale and larger.

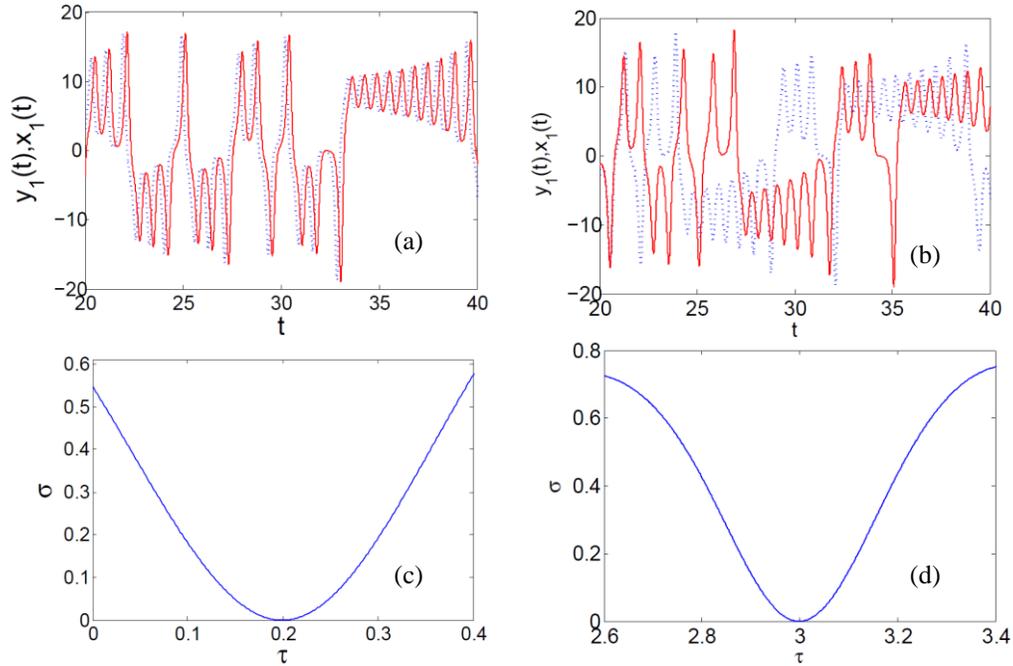

**Fig.3** (color online). Lag synchronization in Lorenz-like 4D systems. Pair of time series of the driver ($y_1$) and response ($x_1$) in dotted (blue) and solid (red) lines respectively shows LS for two different lag time, (a) $\tau=0.2$, (b) $\tau=3.0$. The similarity function plot confirms global minima ($\sigma_{min}$) at corresponding $\tau=0.2$ (c) and 3.0 (d).

We plot the global minima ($\sigma_{min}$) of the similarity function (8) for different lag time ($\tau/T$) in Fig.4. T is the characteristic time scale of the system. The instantaneous phase of a chaotic system is first estimated using the Hilbert transform[1,2] of the time series of one of the state variables and then T is determined by taking an average of the rate of change in instantaneous phase for a long run. Note that the amplitudes of the coupled oscillators are expected to be strongly correlated in a LS state when the global minima ($\sigma_{min}$) is ideally supposed to be zero at the desired lag time. However, in reality, the global minima ($\sigma_{min}$) is seen bounded to very low values ($\sim 10^{-3}$) in Fig.4 even for large lag time which reveals that the error in amplitude correlation is quite small, i.e. the amplitude correlation between the driver and response is very good. The chosen lag time is clearly larger than T and the error is still bounded for both the 3D and 4D systems.

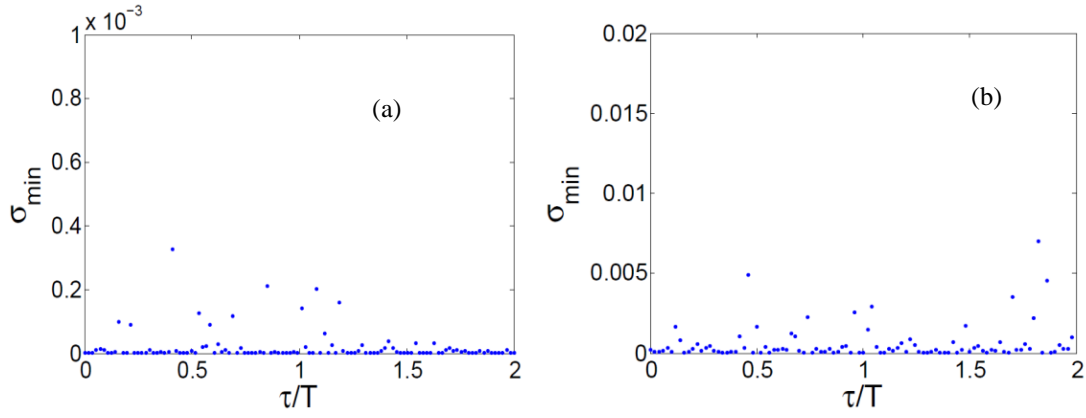

**Fig.4.** Amplitude correlation error. Global minima ($\sigma_{min}$) of the similarity measure is plotted with normalized lag time as an error in amplitude correlation at LS state for (a) 3D Rössler system, (b) 4D Lorenz-like system.

## IV. LAG SYNCHRONIZATION: EXPERIMENT

For physical realization of the OPCL delayed coupling in electronic circuit, we consider a Sprott system[15],

$$\dot{x}_1 = -ax_2, \quad \dot{x}_2 = x_1 + x_3, \quad (14)$$
$$\dot{x}_3 = x_1 + x_2^2 - x_3.$$

A second Sprott system with a parameter mismatch is taken as a driver

$$\dot{y}_1 = -ay_2 - \Delta a y_2, \quad \dot{y}_2 = y_1 + y_3, \quad (15)$$
$$\dot{y}_3 = y_1 + y_2^2 - y_3$$

After coupling, the response (14) becomes

$$\dot{x}_1 = -ax_2 - \alpha \Delta a y_2, \quad \dot{x}_2 = x_1 + x_3, \quad (16)$$
$$\dot{x}_3 = x_1 + x_2^2 - x_3 + \alpha(1-\alpha) y_2^2 + (p - 2\alpha y_{2\tau})(x_2 - \alpha y_{2\tau})$$

The driver and the response are chaotic before coupling for ($a=0.2$, $\Delta a=0.11$). The Hurwitz matrix is constructed from the *Jacobian* of the response details of which is given in ref.11. The coupled response oscillator is given by (16). The circuits for the delay coupled Sprott systems are designed as shown in Fig.5. The driver Sprott circuit (OS-1) is designed using three Op-amps (U1-U3) as integrators with associated resistances, capacitors and an inverting amplifier (U4); the multiplier U5 designs the quadratic nonlinearity in the driver. Similarly the response system (OS-2) is designed using three integrators (U6-U8), one inverting amplifier (U9) and a multiplier U10. The OPCL delay coupler is designed using three op-amps (U11-U13), two multipliers (U14-U15) and a delay line. The delay line in the coupling is designed using a ladder LC network preceded by an isolating amplifier (U16) and followed by a non-inverting amplifier (U17). The non-inverting amplifier U17 is used to compensate the attenuation in the signal due to leakage resistance of the inductors of the LC arrays. A desired lag time is obtained by algebraically adding one after another LC circuit in series. Power supply for all active devices is ±9 Volt. The driver oscillator (OS-1) is connected unidirectionally to the response oscillator (OS-2) via the three terminals (A, B and C). The delay line is inserted between the terminals A and D.

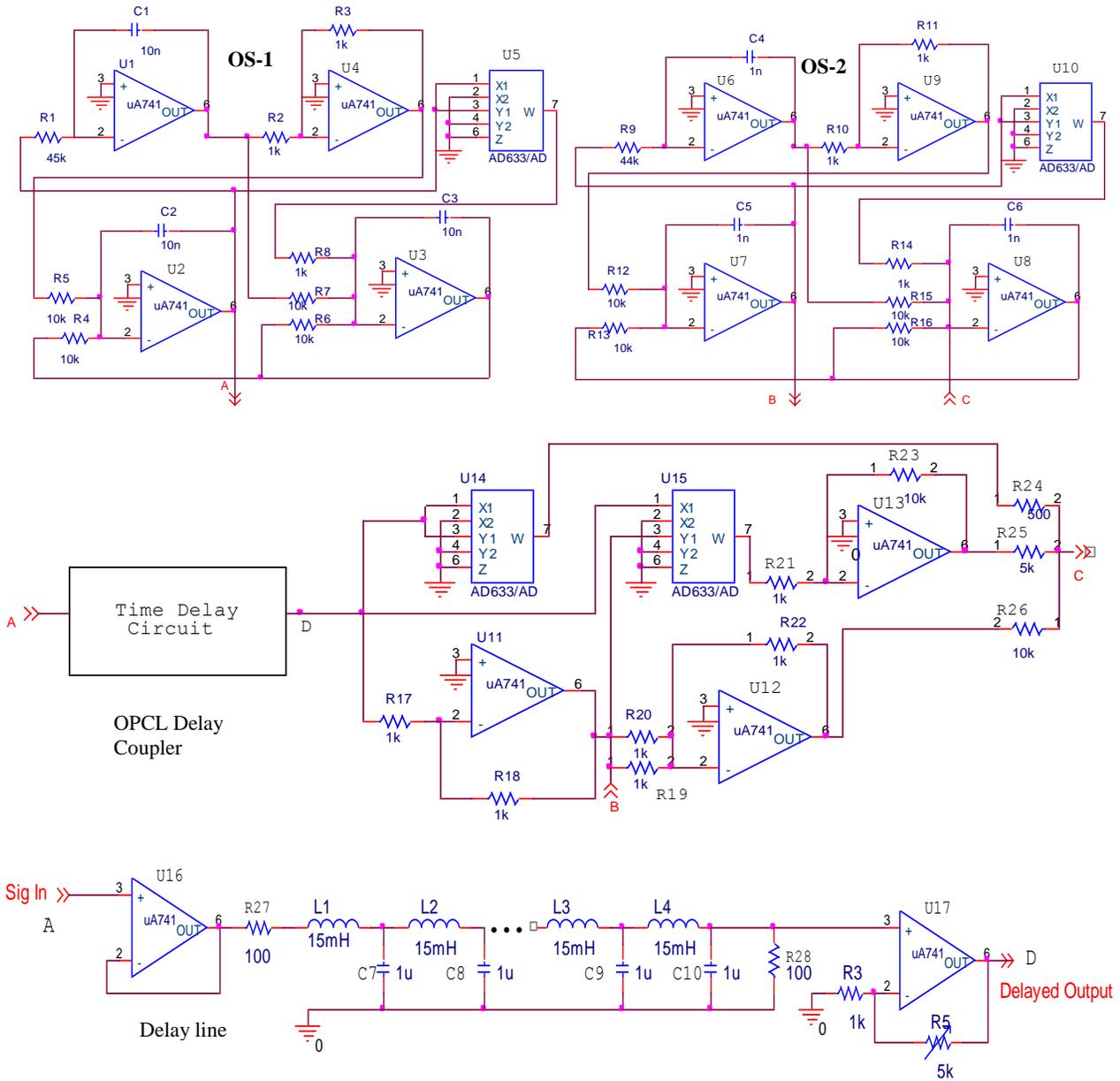

**Fig.5.** Electronic circuit for a delay coupled Sprott oscillator: a driver (OS-1), a response (OS-2) and the OPCL delay coupler. Coupling for the Sprott circuit for α=1 when response signal is not amplified/attenuated. For amplification (α=2), the resistance values in the coupler are chosen as $R_{24}$=166 Ω, $R_{25}$=2.5 kΩ, $R_{18}$=4 kΩ and for attunuation (α=0.5), they are chosen as $R_{24}$=312 Ω, $R_{25}$=10 kΩ and $R_{18}$=500 Ω while other resistance are remaining same shown in the circuit daigram.

Before running the experimental circuit, we first numerically simulate the coupled model (15)-(16) for two desired time lag ($\tau$=2, 8) with three different $\alpha$-values (0.5, 1.0, 2.0). This produces the desired LS effect in the coupled Sprott system with attenuated, identical and attenuated responses respectively. To our best knowledge, we report amplification or attenuation of chaotic signal with a LS scenario for the first time in experiment. Numerical results vis-à-vis experiments are presented in Fig.6.

Numerically simulated state variables $y_1(t)$ and $x_1(t)$ of (15) and (16) are shown in Fig.6(a)-(b). A similar LS scenario is recorded as output voltages of U1 and U6 respectively using a 2-channel digital oscilloscope (Tektronix TDS 2012B, 100MHz, 1GS/s). The oscilloscope pictures of similar driver and response variables are shown in right panels and they are clearly in LS for two desired lag time or delay ($\tau$=150μs, 300μs) as designed by adding one after another LC circuit in the coupler.

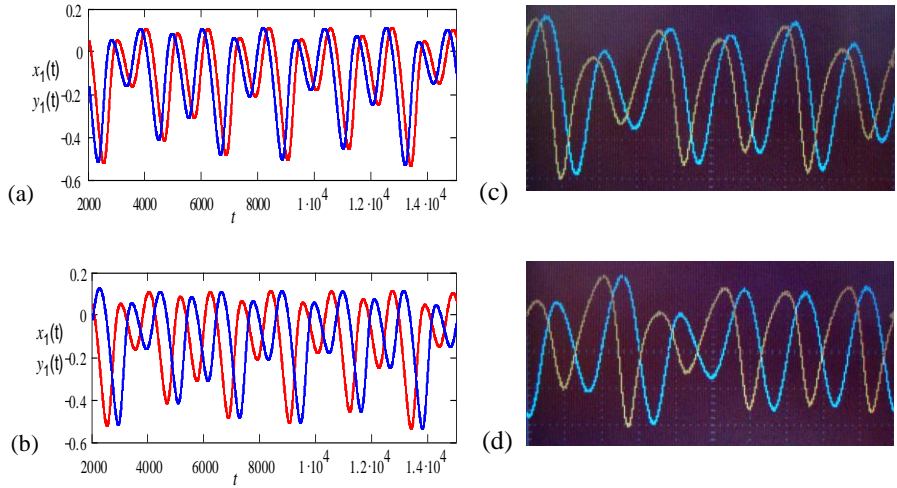

**Fig.6** (color online). Lag synchronization in mismatched Sprott systems. Time series of the driver $y_1(t)$ (red line) and the response $x_1(t)$ (blue line) are plotted for two time lags, $\tau=2$ (a) and $\tau=8$ (b); lag is in arbitrary unit and $\alpha=1$. Hurwitz matrix for Sprott system, H= [0 -1 0; 1 0 1; 1 p -1]$^T$ where p= -3. Oscilloscope pictures of LS: pair of signals $y_1$ (blue line) and $x_1$ (yellow line) with delay time $\tau=150\mu s$ (c), $\tau=300\mu s$ (d). Characteristic time scale or mean time period of the Sprott oscillator: $T=1$ms.

Finally we present experimental evidences of amplification and attenuation in LS state. In previous examples, $\alpha$-values are considered as unity. Now we consider two other values of $\alpha$ (0.5, 2.0). We observe the response signal (blue) is an attenuated version of the chaotic driver signal (red) for $\alpha=0.5$ in Fig. 7(a) while the amplification is seen in Fig.7(b) for amplification of a driver signal (red) at response (blue). Similar experimental scenario is reproduced as shown in the right panels. It may be noted that, in a similar fashion, $\alpha$-values can be chosen negative to induce anti-lag scenarios with amplification or attenuation.

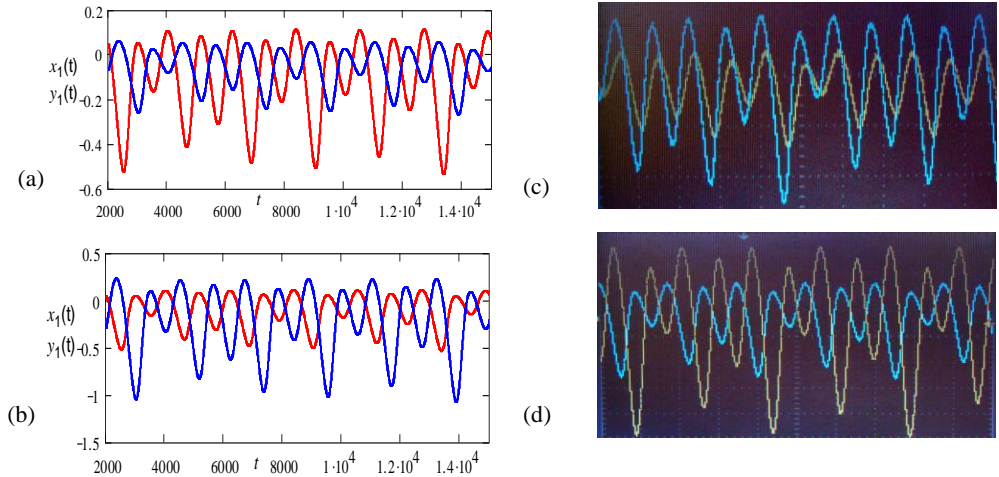

**Fig.7** (color online). Lag synchronization with scaling in mismatched Sprott systems: time series of the driver $y_1$ (red), response $x_1$ (blue) for a time lag $\tau=9$ for (a) $\alpha=0.5$, (b) 2.0. Oscilloscope pictures of the time series of the driver (blue) and response (yellow) for a time lag (c) $\tau=150\mu s$ with attenuation for $\alpha=0.5$, (d) $\tau=450\mu s$ with amplification for $\alpha=2$.

## V. CONCLUSION

We proposed a design of unidirectional delay coupling using an OPCL method based on Hurwitz matrix stability to realize lag synchronization in two chaotic oscillators. The concept of coupling strength is not present in the definition and hence the definition avoids the concept of a critical coupling. We derived the theory of the time-delay coupling to realize stable lag synchronization. We are able to induce any arbitrarily desired time lag between the driver and the response oscillators, identical or mismatched. We showed that error in amplitude correlation is still bounded to a very low value for a lag time even larger than the characteristic time scale of the system. We deliberately inserted a constant scaling parameter in the definition of the coupling to realize amplification and attenuation of the driver attractor at the response oscillator at an LS state. Numerical results are presented in identical Hindmarsh-Rose neuron model, mismatched Rössler oscillator and a Lorenz-like 4D systems. We designed electronic circuits of a Sprott system and experimentally verify all the LS scenarios with amplification and attenuation. Amplification/attenuation of a chaotic signal in a state of lag synchronization, to our best knowledge, is not implemented so far in physical experiment.

**ACKNOWLEDGMENTS**

S.K.B and S.K.D. acknowledge financial support by the BRNS DAE, India (project no.2009/34/26/BRNS).

___________________________________